\documentclass[a4paper]{article}
\usepackage[OT1]{fontenc} 
\usepackage{INTERSPEECH2019}
\usepackage{booktabs}
\usepackage{multirow}
\usepackage{threeparttable}
\usepackage{graphicx}
\usepackage{cite}

\title{A Monaural Speech Enhancement Method for Robust Small-Footprint Keyword Spotting
\thanks{*Corresponding author.}
\thanks{The first two authors contributed equally to this work.}
\thanks{This research was supported in part by the China National Nature Science Foundation (No. 61876214, No. 61866030).}}

\name{Yue Gu$^1$, Zhihao Du$^2$, Hui Zhang$^{1*}$, Xueliang Zhang$^1$}
\address{
$^1$Inner Mongolia Key Laboratory of Mongolian Information Processing Technology, \\
Inner Mongolia University, Hohhot, China\\
$^2$School of Computer Science and Technology, Harbin Institute of Technology, Harbin, China}
\email{427gy@sina.com, 15b903064@hit.edu.cn, alzhu.san@163.com, cszxl@imu.edu.cn}

\begin{document}

\maketitle

\begin{abstract}

Robustness against noise is critical for keyword spotting (KWS) in real-world environments. To improve the robustness, a speech enhancement front-end is involved.  Instead of treating the speech enhancement as a separated preprocessing before the KWS system, in this study, a pre-trained speech enhancement front-end and a convolutional neural networks (CNNs) based KWS system are concatenated, where a feature transformation block is used to transform the output from the enhancement front-end into the KWS system's input. The whole model is trained jointly, thus the linguistic and other useful information from the KWS system can be back-propagated to the enhancement front-end to improve its performance. To fit the small-footprint device, a novel convolution recurrent network is proposed, which needs fewer parameters and computation and does not degrade performance. Furthermore, by changing the input features from the power spectrogram to Mel-spectrogram, less computation and better performance are obtained. our experimental results demonstrate that the proposed method significantly improves the KWS system with respect to noise robustness.

%Robustness against noise is critical for keyword spotting (KWS) task deployed in real-world environments. In this paper, we propose a monaural speech enhancement method for on-device KWS system. The proposed method is based on deep neural network consisting of a pre-trained front-end, feature transformation blocks and a convolutional neural networks (CNNs). Since the speech front-end and KWS system are jointly optimized, the linguistic information contained in the KWS system can be back-propagated to influence the enhancement front-end. To serve the small-footprint purpose, we propose a novel convolution recurrent network which needs less parameters and multiplies without the degradation of performance. We also apply the proposed enhancement models on the Mel-spectrogram domain, which leads to less computation and better performance.With the proposed small-footprint enhancement methods, the resource-limited KWS system can be significant improved with respect to noise robustness.
  
\end{abstract}
\noindent\textbf{Index Terms}: Small footprint, speech enhancement, robust KWS

\section{Introduction}

Keyword spotting (KWS), also called keyword detection (KWD) or spoken term detection (STD), is a crucial technique for human-computer interaction interface. For example, wake-up word detection on mobile devices is an typical scenario. It detects predefined wake-up words in a continuous audio stream. A good KWS system should have low false rejection rate and also low false alarm rate. Moreover, KWS usually runs in ``always-on" mode which requires low power consumption especially in small-footprint embedded systems. 

Currently, the KWS system performs well in a relatively quiet environment. For example, the latest research \cite{Tang2018} achieved an accuracy of 95\% on the Google's Speech Commands Dataset \cite{Warden2018}, with a small model. While, in noisy environments, KWS is still a challenge. In recent years, %the multi-condition training is introduced to improve noise robustness of KWS system in noisy conditions. 
to increase the robustness against noise, a commonly and widely used method is multi-condition training \cite{shan2018attention, Prabhavalkar2015, sainath2015convolutional} which train model with noisy utterances, directly. However, to achieve a good performance, multi-condition training always need a large model, which is impossible to be deployed on devices with limited resources \cite{Yu2018}. 

Recently, with the rise of the deep learning, speech enhancement technique has made a significant progress \cite{wang2018supervised}. 
In the automatic speech recognition (ASR) community, the front-end enhancement techniques have been introduced and have improved the robust ASR systems, where an enhancement front-end is employed to enhance the noisy speech before recognition. Then the recognizer can be trained on clean speech \cite{du2014robust}, or trained on enhanced speech \cite{seltzer2013investigation}.
% The front-end enhancement methods filter out the noise signals from the target speech stream before passing it to KWS system. Recently, speech enhancement has been reformulated as a supervised learning task. Thanks to the rise of the deep learning, speech enhancement has made a significant progress \cite{wang2018supervised}.
% front-end processing which performs acoustic modeling on clean speech. At test phase, the enhancement model is employed to enhance the noisy speech before recognition \cite{du2014robust}. In the retraining strategy, the acoustic model is trained on the enhanced speech at training stage. At test phase, the noisy speech are first enhanced and then fed to the ASR system 
% The third strategy is joint-training in which the enhancement and acoustic model are concatenated to form a larger and deeper network. The network is jointly optimized with the cross entropy \cite{wang2016joint}.
After ASR, the front-end enhancement techniques have also been introduced in KWS. In \cite{Yu2018}, a text-dependent enhancement and KWS method has been developed and shown improvements on the noise robustness. However, its enhancement model is based on bidirectional long-short time memory ( BiLSTM) which needs too many parameters and computation, which does not fit the small-footprint device.
% In \cite{Shan2018}, an attention-based model was developed and achieved superior performance on the single-keyword task under the real noisy environment.

% Based on the performance of multi-condition training, how to further improve the robustness of the model? The front-end enhancement is used to separate interference signals from the noisy speech before passing it for KWS.The Combination of multi-condition training and front-end enhancement is able to significantly improve the robustness of the KWS system.

In this paper, we propose a small-footprint enhancement method for the resource-limited KWS. Compared with the BiLSTM-based models, the proposed model achieves comparable or even better performance with much less parameters and computation.
Considering speech enhancement and keyword spotting are not two independent tasks, they can benefit each other. We concatenate them to build a larger and deeper model, and then optimize them jointly to  improve the noise robustness furtherly. 

Experimental results demonstrate the proposed joint-training method not only significantly outperforms the multi-conditional training method, but also outperforms the enhancement front-end methods, whether its KWS recognizer is trained on clean speech or on enhanced speech. With experiments, we find that for KWS task Mel-spectrogram is a better feature than the power spectrogram, which leads to better performance and lower computation complexity. We also find with Mel-spectrogram the KWS system is less sensitive to the number of phonetic symbols in the keywords.

% we aim at further improving the robustness of KWS system, meanwhile paying more attention to reduce the number of model parameters and the number of model multiplies. Firstly, our work is inspired by the recent success of speech enhancement strategy in speech recognition. The common strategies for KWS task are SE+KWS, Re-training+KWS which performs effective to train a robust system. In our work, we incorporate speech enhancement into end-to-end keyword spotting system, and jointly optimize the two modules both of which are based on deep learning. The proposed method is performs significantly better than the methods based on the above two strategies. Speech enhancement and keyword spotting are not two independent tasks and they can clearly benefit each other. To improve our approach, we further explore the enhancement architectures, including LSTM, BiLSTM and CRN, and compare the performance and footprint of several different network. Then, we find that the enhanced features lost some high frequency information in the process of transforming the Power domain features to the MFCC. To solve this problem, we use the mel domain feature to train the enhancement model so that the enhancement feature can be transformed into mfcc features without any loss.The result of experiment shows that the proposed method outperforms the others methods based on the above two strategies and the recent proposed KWS method \cite{Yu2018}. In addition, our speech enhancement model is nice small-footprint method which achieve the same performance as a complex enhancement model.

\begin{figure}[t]
	\centering
	\includegraphics[width=7cm]{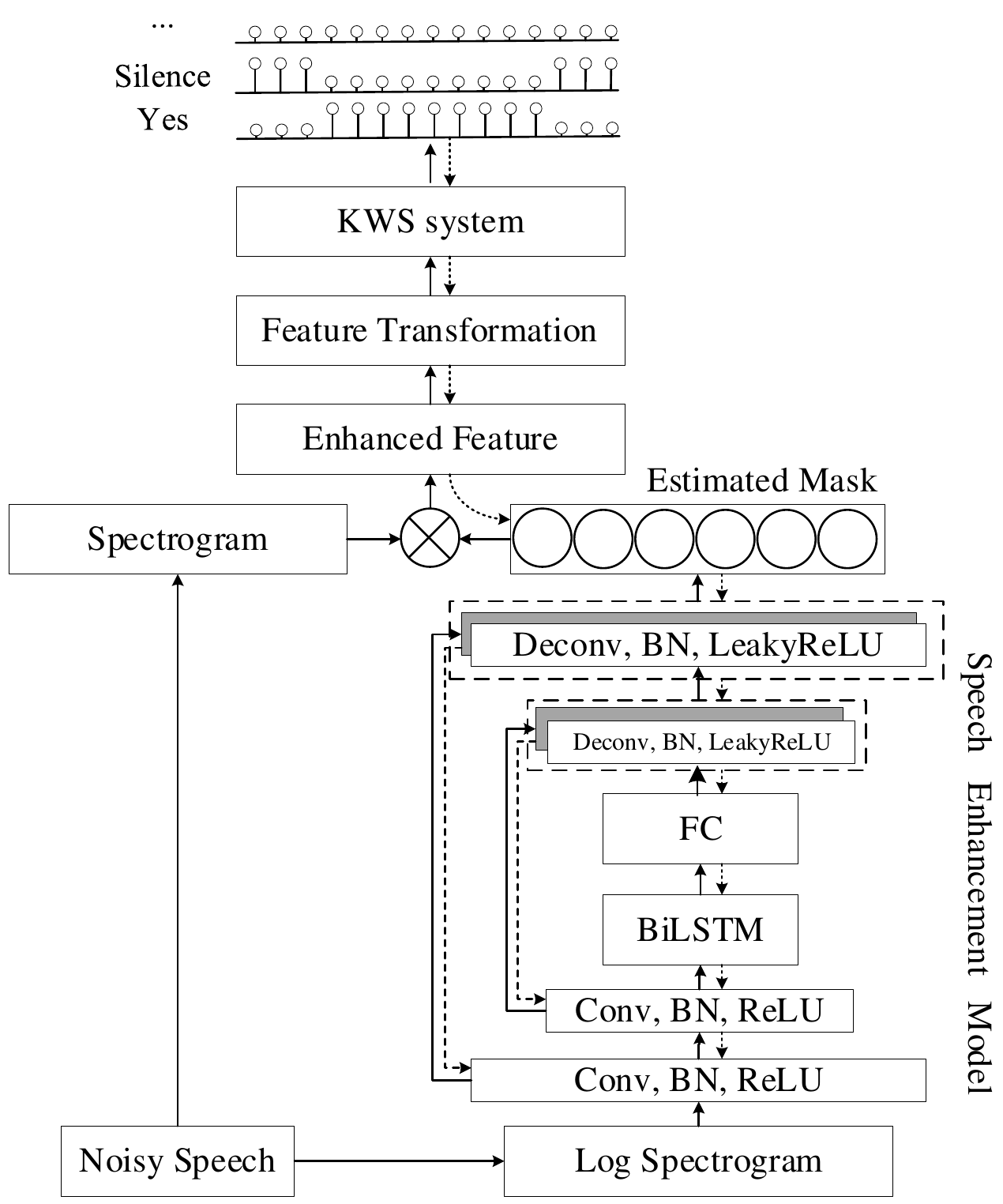}
	\caption{Schematic diagram of the proposed system. Solid and dotted arrows indicate the directions of forward pass and backward pass, respectively. See text for more details.}
	\label{fig:system_framework}
\end{figure}

\section{System description}
The overall framework of our system is shown in Fig. \ref{fig:system_framework}. There are three components in the proposed system, i.e., speech enhancement model, feature transformation block and keyword spotting (KWS) model. The speech enhancement model is trained to predict the ideal ratio masks (IRMs) \cite{narayanan2013ideal}. The enhanced spectrogram are obtained by point-wisely multiplying the noisy spectrogram with the predicted masks. Then the enhanced spectrogram are transformed to the Mel-frequency cepstral coefficients (MFCCs) by the feature transformation block. Given the MFCCs, the KWS model is trained to predict the posterior probability of keywords. The details of these three components are given followingly.

\subsection{Speech enhancement model}
We employ the masking-based speech enhancement method, which has successfully improved the human speech perceptive quality \cite{wang2014training} and the noise robustness of ASR \cite{wang2016joint}. The loss function of masking-based method is defined as:
\begin{equation}
\mathcal{L}_{\mathbf{MSE}} = \frac{1}{T}\frac{1}{F}\sum_{t=1}^{T}\sum_{f=1}^{F}{\lVert M(t, f) - \hat{M}(t,f)\rVert_2^2} \label{eq:mes_loss}
\end{equation}
where $M(t, f)$ and $\hat{M}(t,f)$ are the ideal and predicted time-frequency (T-F) mask at time $t$ and frequency $f$, respectively. $T$ and $F$ are the total number of frames and frequency bins respectively. The IRM $M$ is defined as:
\begin{equation}
M(t, f) = \sqrt{\frac{|S(t, f)|^2}{|S(t, f)|^2 + |N(t, f)|^2}}
\end{equation}
% where $M$ is the IRM of a noisy keyword created by mixing a noise-free keyword with a noise signal at a specific SNR level, 
where $S$ represents the spectrogram of the clean speech, $N$ stands for the spectrogram of the noise signal.
% where $S(t, f)$ and $N(t, f)$ are the T-F units of clean speech and noise at time $t$ and frequency $f$, respectively.
% $M(t, f)$ is the ideal mask at time $t$ and frequency $f$.
% The ideal mask belongs to $[0, 1]$. 

In test stage, the IRM is predicted from the noisy speech and the enhanced spectrogram can be obtained by:
\begin{equation}
\hat{S} = Y \otimes \hat{M} \label{eq:enhanced_feats}
\end{equation}
where $\hat{M}$ is the mask predicted by the enhancement model. $Y$ is the spectrogram of the noisy signal. $\otimes$ represents the element-wise matrix multiplication.

The IRM can be defined in different T-F domains.
Although the power spectrogram is a common choice in the speech enhancement community, there are better choice. In the proposed KWS system, the output of the enhancement model is feed into KWS model which requires the MFCC as input. While, the frequency bins in the power spectrograms are integrated to extract MFCCs by the Mel-filter bank. It means many information contained in the power spectrogram are filtered out. Therefore, it is not efficient and necessary to perform enhancement on the power spectrogram. In contrast, the Mel-spectrogram can be used to extract MFCCs, directly. So that we use the proposed enhancement model to predict IRM on the Mel-spectrogram. In this way, the spectrogram of noise $N$, clean speech $S$, and noisy speech $Y$, are all in the form of Mel-spectrogram.

% The proposed enhancement method is performed on the Mel-spectrogram domain.
% We find it is a more efficient choice to perform the enhancement on the Mel-spectrogram domain. 
% In this paper, the power and Mel-spectrogram domain are employed to perform the enhancement. 
% For the power spectrogram domain, two enhancement models are evaluated. One is the conventional bidirectional recurrent neural network (RNN) with LSTM cells, The other is the convolution recurrent networks (CRN) \cite{tan2018convolutional}. Inspired by the success of CRNs on improving speech intelligibility and quality \cite{tan2018convolutional}, we design the novel small-footprint CRNs with the limiting parameters and multiplies.

To serve the small-footprint purpose, we design a novel convolution recurrent network (CRN) with the limiting parameters and computation. 
The architecture of CRN is shown as speech enhancement model in the lower part of the Fig. \ref{fig:system_framework}. There are two components in the CRN, i.e., the convolutional encoder-decoder and the RNN with LSTM cells followed by a linear projection layer. Skip connections are added to the corresponding layers between the encoder and decoder. Batch normalization \cite{ioffe2015batch} and rectified linear units (ReLUs) \cite{nair2010rectified} are employed in the convolutional layers and the leaky ReLUs (lReLUs) are used in the de-convolutional layers instead of ReLUs. Sigmoid nonlinearity is employed for the output layer. 
% CRNs are originally developed in \cite{tan2018convolutional}.
Note that there are two differences between the CRNs in \cite{tan2018convolutional} and ours.
Firstly, with the limiting parameters and computation, the convolution layers in our CRNs have the strides on both time and frequency axises while the origin CRN only strides on the frequency axis. Secondly, we employ the lReLU at the decoding stage, which guarantees the nonzero gradients everywhere and benefits the optimizing processing of the encoding stage.

% The enhancement model can be trained on the power or Mel-spectrogram domains. 
\subsection{Feature transformation block}
\begin{figure}[ht]
	\caption{The feature transformation block for (a) Mel-spectrogram and (b) power spectrogram.}
	\label{fig:ftblock}
	\begin{minipage}[b]{0.45\linewidth}
		\centering
		\centerline{\includegraphics[width=3.2cm]{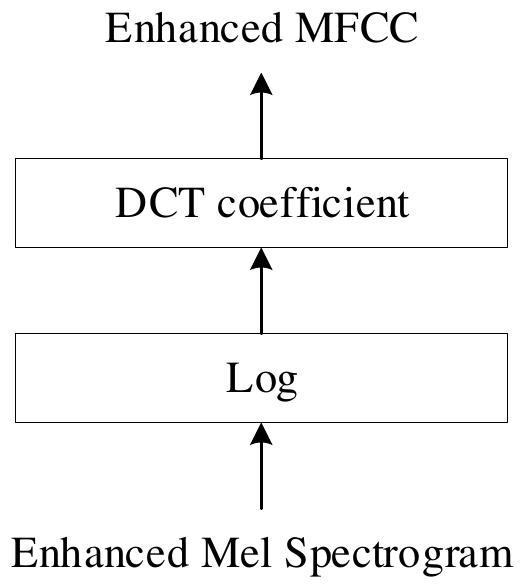}}
		%  \vspace{1.5cm}
		\centerline{(a) Mel-spectrogram to MFCC}
	\end{minipage}
	\hfill
	\begin{minipage}[b]{0.45\linewidth}
		\centering
		\centerline{\includegraphics[width=3.4cm]{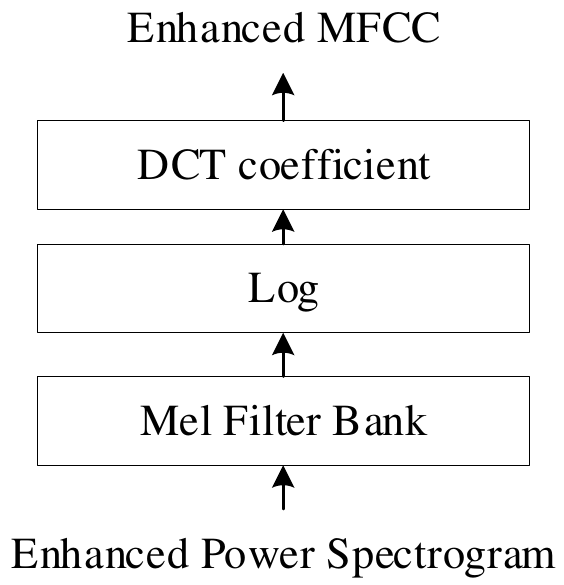}}
		%  \vspace{1.5cm}
		\centerline{(b) Power spectrogram to MFCC}
	\end{minipage}
\end{figure}

The input of KWS system is MFCC while the outputs of enhancement model are spectrograms. To extract the MFCCs from the spectrograms, we design the feature transformation blocks (FTBs) which are shown in Fig. \ref{fig:ftblock}. Transforming Mel-spectrogram to MFCC needs taking logarithm firstly then applying discrete cosine transformation (DCT). For comparison, an enhancement model trained to predict the IRM on the power spectrogram is also built. For this model, we need transform the power spectrogram into MFCC. Similar to the Mel-spectrogram, to obtain MFCC from power spectrogram, the input should pass a Mel-filter bank, then take logarithm, at last apply a DCT. Note that both the Mel-filter bank filtering and the DCT can be implemented with the matrix multiplication which can be further represented as the linear layers in a neural network \cite{sainath2013learning}. As a result, included the FTBs, the proposed systems can be trained with back-propagation algorithm.

% The enhanced power spectrograms are first multiplied with Mel-filter bank to obtain the Mel-spectrograms, and then compressed by logarithm, finally, they are transformed by discrete cosine transformation (DCT) to obtain MFCC. Similar, the enhanced Mel-spectrograms are first compressed by logarithm and then transformed by DCT. Note that both the filtering with Mel-filter bank and the DCT can be implemented with the matrix multiplication which can be further represented as the linear layers in a neural network \cite{sainath2013learning}.
\subsection{Keyword spotting system}
We employ the model \texttt{cnn-trad-pool2} developed in \cite{Tang2017} as our KWS system. This model diverges slightly from the model \texttt{cnn-trad-fpool3} which is originally introduced in \cite{tan2018convolutional}. The size and stride of the first max-pooling layer are set to $(2, 2)$ and the hidden linear layers are dropped in \texttt{cnn-trad-pool2}, which leads to better accuracy. % The details of \texttt{cnn-trad-pool2} can be found in \cite{Tang2017}

\section{Experiments and results}

\subsection{Experimental settings}
We evaluated the proposed models on Google's Speech Commands Dataset which contains 105,829 one-second long utterances and 6 background noise records (including pink noise, white noise, and daily environmental sounds such as doing the dishes, exercise bike, etc.) \cite{Warden2018}.
% which contains 105,829 one-second long utterances of 35 short words recorded by thousands of difference people, as well as background noise records such as pink noise, white noise, and daily environmental sounds which include kitchen, exercise bike, running and miaowing \cite{Warden2018}.
% The paper announcing the data release also references Google's TensorFlow implementation of Sainath and Parada's models, which provide the basis of our comparison. 
Following Google's implementation, the task is to detect 10 keywords, unknown and silence.
% :"yes", "no", "up", "down", "left", "right", "on", "off", "stop", "go", unknown, or silence.
In our experiments, the baseline \texttt{cnn-trad-pool2} follows the exactly the same procedure as the TensorFlow reference.
% implemented by Tang et al. \cite{Tang2017}.
% as the TensorFlow reference, the implementation is referenced by Tang et al. \cite{Tang2017}. 
The dataset is split into the training, validation, and test set with the ratio of 8:1:1. Noisy utterances are obtained by mixing up with 6 noises at signal-to-noise ratios (SNRs) of $\{-3, 0 , 3, 6\}$. There are roughly 812k noisy examples for training and 97.6k each for validation and test.
% Each noise in the dataset is about one minutes.
% , randomly select one second from it, then mix it up with a one second keyword utterance. After mixing noise, the amount of match-noise condition data is expanded to 24 times before, roughly 812k examples for training and 97.6k each for validation and testing. 
% In order to test our trained models more fairly and accurately, 
Another 25 keywords are employed to evaluate the models, which are not involved at the training phase.
% We mix noises with these 25 words which have not been seen in train set. 
Finally, the test set contains 210k noisy utterances with keywords and non-keywords ratio of 1.3:1. To evaluate the generalization of the models, 100 Nonspeech Sounds \footnote{http://web.cse.ohio-state.edu/pnl/corpus/HuNonspeech/HuCorpus.html} are employed, which are unseen at the training stage. The unmatched test set contains nearly 3.6M utterances.
All the utterances are sampled to 16~kHz and the features are extracted with the window length of 30~ms and the shift length of 10~ms. The 480-point short-time Fourier transform is employed. The Mel-filter bank is calculated with the low frequency 20~Hz and high frequency 4~KHz. The 40-dimension DCT coefficients are used to extract MFCC.

Accuracy is the main metric, which is simply measured as the fraction of classification decisions that correct. We also plot receiver operating characteristic (ROC) curves, where the $x$ and $y$ axes show false alarm rate (FAR) and false reject rate (FRR), respectively.
% For a given sensitivity threshold defined as the minimum probability at which a class is considered positive during evaluation, FAR and FRR represent the probabilities of obtaining false positives and negatives, respectively. By sweeping the sensitivity interval $[0.0, 1.0]$, curves for each of the keywords are computed and then averaged vertically to produce the overall curve for a specific model. 
Methods with less area under the curve (AUC) are better. Equal error rates (EERs) are also employed to shows the KWS performance with the enhancement models.
% coincides with the objective evaluation of the fronted speech quality.

All models are trained with the Adam optimizer \cite{Kingma2014} and the mini-batch size of 256 on the utterance-level. We set the learning rate to 0.0001. The mean squared error (MSE) and cross entropy (CE) are the objective functions of the enhancement model and KWS system, respectively. 
% We train the whole model with a minibatch size of 256.
The best models are selected by the best accuracy on the validation set.
%tnote[1]{https://storage.cloud.google.com/download.tensorflow.org/data/speech_commands_v0.02.tar.gz}.

We evaluate the proposed small-footprint CRNs on the power and Mel-spectrogram. For each spectrogram, we design two models with different model size. We refer the full-size model trained on the power and Mel-spectrogram as \texttt{PowCRN32} and \texttt{MelCRN32} respectively, and the narrow models are referred as \texttt{PowCRN16} and \texttt{MelCRN16} respectively. The details of CRNs are shown in Tab. \ref{tab:CRN}. As the comparison, a LSTM-based model is also evaluated, which consists of two hidden layers with 384 bidirectional LSTM cells followed by a linear projection layer with 241 units. We refer this enhancement model as \texttt{BiLSTM} \cite{Yu2018}. The model size is given in Tab. \ref{tab:para_multi}. In Tab. \ref{tab:para_multi}, we list the parameter numbers and the computation complexity evaluated by the number of multiply operation for each model.

\begin{table}[h]
	\scriptsize
	\caption{The architectures of small-footprint CRNs. $T$ denotes the number of time frames in the spectrogram. $(f, h)$ is set to $(16, 32)$ and $(32, 64)$ for the narrow and full-size CRNs, respectively. For convolution and deconvolution layers, the parameter indicates kernel size, stride and filter number. $h$ stands for the number of bidirectional LSTM cells.}
	\label{tab:CRN}
	\centering
	\renewcommand\arraystretch{1.3}
	\renewcommand\tabcolsep{3.0pt}
	\begin{tabular}{c|c|c|c}
		\toprule
		\textbf{Layer Name} & \textbf{Input Size} & \textbf{Parameter} & \textbf{Output Size}\\
		\midrule
		\multicolumn{4}{c}{PowCRN} \\
		\hline
		reshape\_1 & $T \times 241 $ & - & $ 1 \times T \times 241 $\\
		conv\_1 & $ 1 \times T \times 241 $ & $ 8, 4, f $ & $ f \times T/4 \times 60 $ \\
		conv\_2 & $ f \times T/4 \times 60 $ & $ 8, 4, f $ & $ f \times T/16 \times 15 $ \\
		reshape\_1 & $ f \times T/16 \times 15 $ & - & $ T/16 \times 15f $\\
		BiLSTM & $ T/16 \times 15f $ & $h$ & $ T/16 \times h $\\
		FC & $ T/16 \times h $ & $ 15f $ & $ T/16 \times 15f $ \\ 
		reshape\_2 & $ T/16 \times 15f $ & - & $ 2f \times T/16 \times 15 $ \\
		deconv\_2 & $ 2f \times T/16 \times 15 $ & $ 8, 4, f $ & $ 2f \times T/4 \times 60 $ \\
		deconv\_1 & $ 2f \times T/4 \times 60 $ & $ 9, 4, f $ & $f \times T \times 241 $ \\
		conv\_out & $f \times T \times 241 $ & $ 3, 1, 1 $ & $ 1 \times T \times 241 $ \\
		reshape\_3 & $ 1 \times T \times 241 $ & - & $ T \times 241 $ \\
		\hline
		\multicolumn{4}{c}{MelCRN} \\
		\hline
		reshape\_1 & $T \times 40 $ & - & $ 1 \times T \times 40 $\\
		conv\_1 & $ 1 \times T \times 40 $ & $ 4, 2, f $ & $ f \times T/2 \times 20 $ \\
		conv\_2 & $ f \times T/2 \times 20 $ & $ 4, 2, 2f $ & $ 2f \times T/4 \times 10 $ \\
		conv\_3 & $ 2f \times T/4 \times 10 $ & $ (3, 4), (1, 2), 4f$ & $ 4f \times T/4 \times 5 $ \\
		reshape\_1 & $ 4f \times T/4 \times 5 $ & - & $ T/4 \times 20f $ \\
		BiLSTM & $ T/4 \times 20f $ & $h$ & $ T/4 \times h $ \\
		FC & $ T/4 \times h $ & $ 20f $ & $ T/4 \times 20f $ \\ 
		reshape\_2 & $ T/4 \times 20f $ & - & $ 8f \times T/4 \times 5 $ \\
		deconv\_3 & $ 8f \times T/4 \times 5 $ & $ (3,4),  (1,2), 2f $ & $ 4f \times T/4 \times 10 $ \\
		deconv\_2 & $ 4f \times T/4 \times 10 $ & $ 4, 2, f $ & $ 2f \times T/2 \times 20 $ \\
		deconv\_1 & $ 2f \times T/2 \times 20 $ & $ 4, 2, f $ & $ f \times T \times 40 $ \\
		conv\_out & $ f \times T \times 40 $ & $ 3, 1, 1 $ & $ 1 \times T \times 40 $ \\
		reshape\_3 & $ 1 \times T \times 40 $ & - & $ T \times 40 $ \\
		\bottomrule
	\end{tabular}
\end{table}

\begin{table}[h]
	% \footnotesize
	\caption{The number of parameters and multiplies used for the KWS system and different enhancement models.}
	\label{tab:para_multi}
	\centering
	\renewcommand\tabcolsep{13.0pt}
	\begin{tabular}{c|c|c}
		\toprule
		\textbf{Model Name} & \textbf{Parameters} & \textbf{Multiplies}\\
		\midrule
		cnn-trad-pool2 & 493.7K & 95.87M \\
		\hline
		BiLSTM & 5661.0K & 432.7M \\
		PowCRN32 & 724.0K & 280.1M \\
		PowCRN16 & 182.3K & 73.0M \\
		MelCRN32 & 881.3K & 115.1M \\
		MelCRN16 & 221.5K & 29.2M \\
		\bottomrule
	\end{tabular}
\end{table}

Beside the baseline \texttt{cnn-trad-pool2} which uses the multi-conditional training technique, we apply three training strategys for all other enhancement front-end based models. Firstly, the enhancement model is pre-trained against the MSE loss as Equation (\ref{eq:mes_loss}). Then, the enhancement model is concatenated to the KWS model through the feature translation block. In these enhancement front-end based models, KWS model can be trained alone with the MFCC of noisy utterances, which we refer it as \texttt{KWS+\{enhancement model\}}. KWS model also can be trained alone with the MFCC of enhanced spectrogram, which we refer it as \texttt{retrain+\{enhancement model\}}. In fact, KWS model and the enhancement model can be trained together with the noisy spectrogram, which we refer it as \texttt{joint+\{enhancement model\}}.

% The multi-conditional training (MCT) strategy has been employed to improve the noise robustness of KWS \cite{sainath2015convolutional}. We employ the MCT strategy to train our KWS model as the baseline. And we pre-train the enhancement model with the loss function defined in Equation.(\ref{eq:mes_loss}).

% In the independent font-end strategy, the pre-trained enhancement models, KWS system and FTBs are fixed. The noisy keywords are directly fed to the enhancement model. The enhanced features are transformed to MFCC by FTB, and then the MFCCs are fed to the KWS system to perform classification. We refer this strategy as \texttt{KWS+\{enhancement model\}}.

% For the retraining strategy, the pre-trained enhancement models and FTBs are fixed, and the KWS system is retrained with the enhanced features at training stage. At test phase, the retrained KWS model is fed with the enhanced MFCCs which are transformed from the enhanced features by the FTBs. We refer this strategy as \texttt{retrain+\{enhancement model\}}.

% Inspired by the joint-training strategy in the robust ASR, we concatenate the pre-trained enhancement models, FTBs and KWS system to build a larger and deeper network. Note that the FTBs are also trainable. At the training stage, this deeper network are jointly optimized with the cross entropy on the noisy keywords. At the test phase, the noisy keywords are directly fed to the deeper network and the categories of the keywords are predicted. This strategy is referred as \texttt{joint+\{enhancement model\}} in this paper.

\begin{table}[t]
	\footnotesize
	\caption{ The test accuracy, EER and AUR of each model under matched noise condition.}
	\label{tab:test_accuracy}
	\centering
	\renewcommand\arraystretch{1.3}
	\renewcommand\tabcolsep{3.0pt}
		\begin{tabular}{cc|cc}
		\toprule
		\textbf{Model}      & \textbf{Test accuracy(\%)}     & \textbf{AUC (\%)} & \textbf{EER (\%)}           \\
		\midrule
		cnn-trad-pool2            & 80.89                      & 1.99 & 7.28          \\
		\hline
		KWS+BiLSTM                & 87.64                     & 1.30 & 6.66          \\
		retrain+BiLSTM           & 90.18                      & 1.17 & \textbf{5.92}          \\
		joint+BiLSTM        & 91.64                & \textbf{1.01}  & 6.15        \\
		\hline
		KWS+PowCRN32          & 86.42                      &1.52  &   6.67      \\
		retrain+PowCRN32     &      87.69                & 1.53 &    6.63      \\
		joint+PowCRN32  &      91.07              & 1.20 &     6.27      \\
		\hline
		KWS+PowCRN16          & 86.20                      & 1.61 & 6.73          \\
		retrain+PowCRN16     &     87.01            & 1.67 &    6.88       \\
		joint+PowCRN16  & 90.68                      & 1.22  & 6.50         \\
		\hline
		KWS+MelCRN32          & 87.59                    & 1.59 &  6.97       \\
		retrain+MelCRN32     & 89.17                    & 1.35 &   6.10       \\
		joint+MelCRN32  & \textbf{93.17}                    & 1.19 & 6.20          \\
		\hline
		KWS+MelCRN16          & 86.87                      & 1.64 & 7.00         \\
		retrain+MelCRN16     &    88.20                  & 1.42 &      6.49   \\
		joint+MelCRN16  & 92.56                     & 1.28 & 6.39          \\
		\bottomrule
	\end{tabular}
\end{table}

\begin{table}[t]
	\footnotesize
	\caption{The test accuracy of joint-trianing models under unmatched noise condition.}
	\label{tab:test_accuracy_unmatched}
	\centering
	\renewcommand\arraystretch{1.3}
	\renewcommand\tabcolsep{3.0pt}
		\begin{tabular}{cc|cc}
		\toprule
		\textbf{Model}      & \textbf{Accuracy(\%)}    & \textbf{Model}      & \textbf{Accuracy(\%)}           \\
		\midrule
		cnn-trad-pool2            & 68.81                      & joint+BiLSTM  & 73.74          \\
		\hline
		joint+PowCRN32  &      75.19             &joint+PowCRN16 &     72.49      \\
		\hline
		joint+MelCRN32         & \textbf{78.12 }               & joint+MelCRN16  & 75.67       \\
		\bottomrule
	\end{tabular}
\end{table}

\begin{figure}[t]
\caption{ROCs from the perspective of (a) different enhancement models, (b) training strategy, (c) feature domain. And (d) AUC reduction against phonetic symbol length.}
\label{fig:rocs}
	\begin{minipage}[b]{0.45\linewidth}
		\centering
		\centerline{\includegraphics[width=4.2cm]{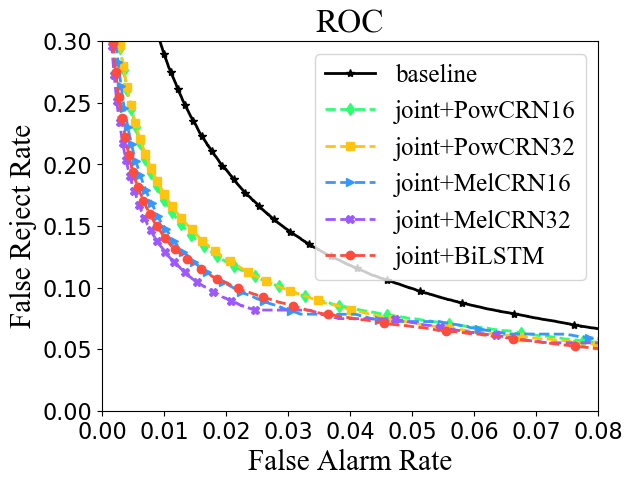}}
		%  \vspace{1.5cm}
		\centerline{(a) Model}
	\end{minipage}
	\hfill
	\begin{minipage}[b]{0.45\linewidth}
		\centering
		\centerline{\includegraphics[width=4.2cm]{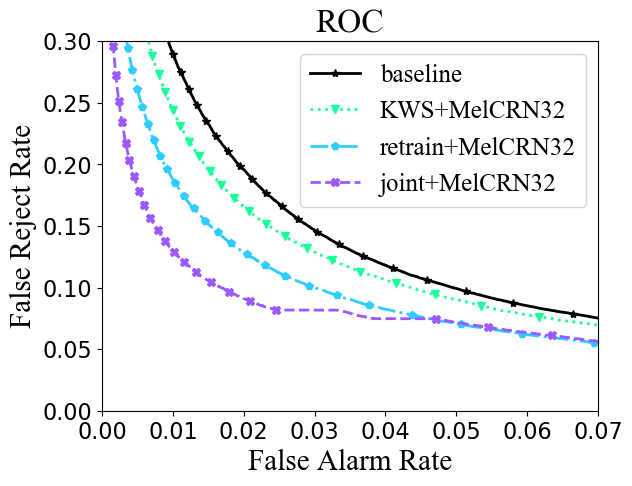}}
		%  \vspace{1.5cm}
		\centerline{(b) Training Strategy}
	\end{minipage}
	\hfill
	\begin{minipage}[b]{0.45\linewidth}
		\centering
		\centerline{\includegraphics[width=4.2cm]{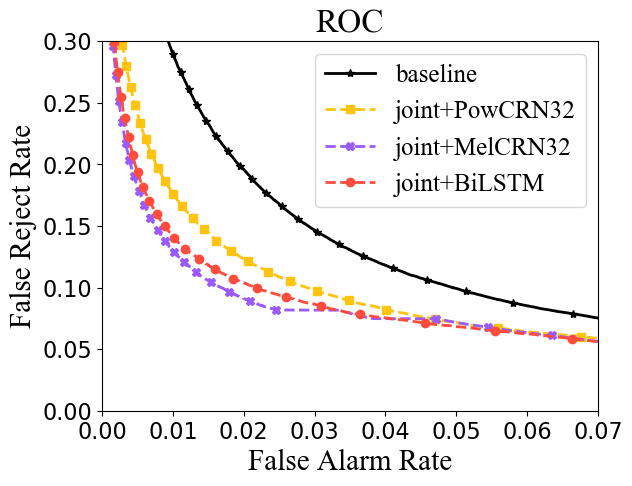}}
		%  \vspace{1.5cm}
		\centerline{(c) Feature Domain}
	\end{minipage}
	\hfill
	\begin{minipage}[b]{0.45\linewidth}
		\centering
		\centerline{\includegraphics[width=4.2cm]{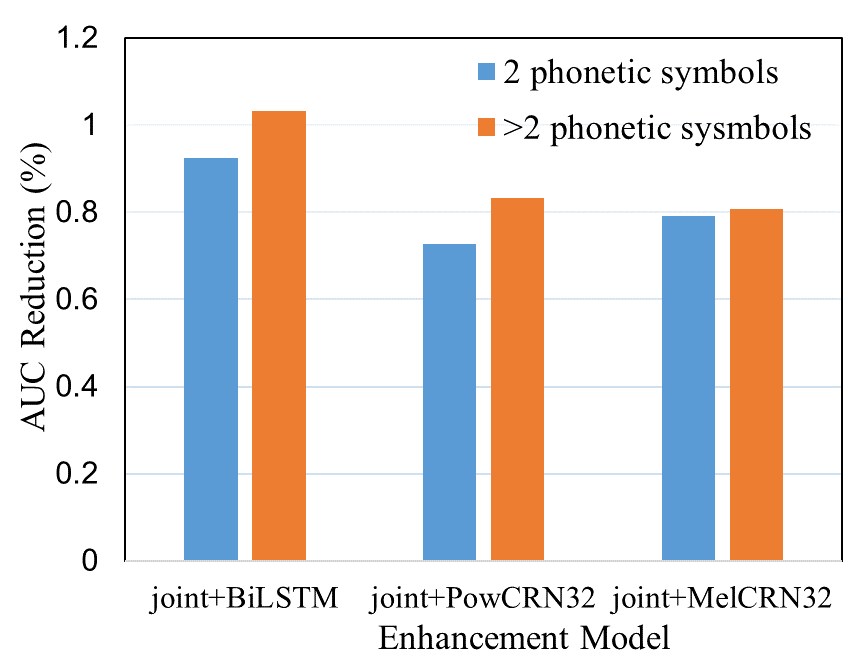}}
		%  \vspace{1.5cm}
		\centerline{(d) AUC Reduction}
	\end{minipage}
\end{figure}

\subsection{Results}
The experimental results are given in Tab. \ref{tab:test_accuracy} and Fig. \ref{fig:rocs}. 

\textbf{Model comparison: } From Tab. \ref{tab:test_accuracy} and Fig. \ref{fig:rocs} (a), we can see all of the comparison models outperform the baseline. The performance of the BiLSTM-based model is good, however its parameter number and computation is the hugest (seen in Tab. \ref{tab:para_multi}) which doesn't serve the small-footprint purpose. The proposed CRNs have acceptable parameters and needs less computation, but have achieved comparable performance compared with BiLSTM-based model. The parameters and the required multiplies are further reduced in the narrow model (PowCRN16, MelCRN16), but it also obtained a comparable performance with the BiLSTM-based model. 
% In addition, to test the generalization ability to new noise, we evaluated the models based on joint-training strategy on a noise-unmatched test set which contains nearly 3,600,000 utterances. Tab.\ref{tab:test_accuracy_unmatched} shows the results 

\textbf{Training strategy: } From Tab. \ref{tab:test_accuracy} and Fig. \ref{fig:rocs} (b), we can see all of the enhancement front-end based models outperform the multi-conditional trained baseline. Specifically, the retrained KWS model trained with enhanced spectrogram is better than the KWS model trained with noisy utterances, and the joint-trained KWS model is better than the retrained KWS model. It is because that the mismatch between the enhancement model and KWS model is descending in the order of model trained with clean utterances, retrained model and joint-trained model. Especially, for the small-footprint enhancement models (PowCRNs and MelCRNs), the joint-training strategy significantly improves the performance.

% From Tab. \ref{tab:test_accuracy} and Fig. \ref{fig:rocs} (a), we find that the joint-training strategy is suitable for all the enhancement model and outperforms the retraining and independent front-end processing strategy with respect to the accuracy and AUC. Especially for the small-footprint enhancement models (PowCRNs and MelCRNs), the joint-training strategy significantly improves the performance.

\textbf{Mel vs power spectrogram: } From Tab. \ref{tab:test_accuracy} and Fig. \ref{fig:rocs} (c), we find the CRNs trained on the Mel-spectrogram have better performance and similar parameters compared with the models trained on the power spectrogram. Since the dimension of Mel-spectrograms is much less than the power spectrograms, the multiplies of the enhancement models can be significantly reduced. We think the Mel-spectrogram is more suitable for the KWS system. Beacuse the input of KWS system is always silence, background noise or non-speech, false alarms on those must be minimized. With the limitation of low FAR ($<2.0\%$), we find \text{MelCRN32} achieves the lowest FRR than the \text{PowCRN32} and \text{BiLSTM}. This advantage is also retained by the narrow model. % MelCRN16 when it is compared with PowCRN16.

% Since the enhanced power spectrograms will be integrated to extract MFCCs by the Mel-filter bank, it is more effective and efficient to perform enhancement on the Mel-spectrogram domains.
% it seems a better choice to perform the speech enhancement on the Mel-spectrogram domain for the small-footprint KWS system. The multiplies of MelCRNs are much less than the BiLSTM and PowCRNs, while the model size of the MelCRNs and PowCRNs are similar. 
\textbf{Sensibility on phonetic symbol length: }
Since the keywords have different phonetic symbols, we wonder whether enhancement models are sensitive to the number of phonetic symbols in the keywords. 
% also explore the relationship between the requirement of enhancement models and the number of phonetic symbols.
We split the dataset into two sets, i.e., the keywords with 2 and more phonetic symbols. Fig. \ref{fig:rocs} (d) shows AUC reductions for keywords with different number of phonetic symbols, where the less reduction the better. From the figure, we can see that the Mel-spectrogram based method is less sensitive to the number of phonetic symbols in the keywords than the models on the power spectrogram.

\textbf{Noise generalization: }
Tab. \ref{tab:test_accuracy_unmatched} shows the results of joint-training models under the unmatched noise condition which contains 100 unseen noises. From the table, we can see the proposed full-size CRNs have better generalization to new noise conditions than the BiLSTM. And the CRNs on Mel spectrogram domain achieves higher accuracy than that on power spectrogram domain.

\section{CONCLUSIONS}

In this paper, we proposed a small-footprint speech enhancement technique for robust KWS, which integrates a front-end enhancement model and a back-end KWS model. 
The proposed CRNs achieve better performance under both matched and unmatched noise condition, and CRNs need less parameters and computation compared with the conventional BiLSTM-based model. We find Mel-spectrogram is better than power spectrogram because it can achieve comparable performance with less computation and similar or smaller model size.
Beside that the Mel-spectrogram based method is non-sensitive to the phonetic symbol length in the keywords.

% the keywords with less phonetic symbols make the keyword spotting difficult and have more requirements of the speech enhancement.
% Compared with TDSE method\cite{Yu2018} and two effective training(SE+KWS, Re-training+KWS) strategy, the systematic evaluation shows that our proposed method achives superior performance in low SNRs condition. we explore the speech enhancement architectures to improve the performance and decrease the footprint of KWS system.Experiments show that BiLSTM is preferred over LSTM ,the best performance is achieved by CRNN and CRNN has the smallest model complexity.We also explore speech enhancement feature representation, our result shows that the Mel-domain feature has a better performance than the Power-domain feature and the former has less model parameters and model multiplies.

%\section{Acknowledgements}

%This research was supported by National Natural Science Foundation of China No.61876214.

\bibliographystyle{IEEEtran}

\bibliography{mybib}

% Generated by IEEEtran.bst, version: 1.13 (2008/09/30)
\begin{thebibliography}{10}
\providecommand{\url}[1]{#1}
\csname url@samestyle\endcsname
\providecommand{\newblock}{\relax}
\providecommand{\bibinfo}[2]{#2}
\providecommand{\BIBentrySTDinterwordspacing}{\spaceskip=0pt\relax}
\providecommand{\BIBentryALTinterwordstretchfactor}{4}
\providecommand{\BIBentryALTinterwordspacing}{\spaceskip=\fontdimen2\font plus
\BIBentryALTinterwordstretchfactor\fontdimen3\font minus
  \fontdimen4\font\relax}
\providecommand{\BIBforeignlanguage}[2]{{%
\expandafter\ifx\csname l@#1\endcsname\relax
\typeout{** WARNING: IEEEtran.bst: No hyphenation pattern has been}%
\typeout{** loaded for the language `#1'. Using the pattern for}%
\typeout{** the default language instead.}%
\else
\language=\csname l@#1\endcsname
\fi
#2}}
\providecommand{\BIBdecl}{\relax}
\BIBdecl

\bibitem{Tang2018}
R.~Tang and J.~Lin, ``Deep residual learning for small-footprint keyword
  spotting,'' pp. 5484--5488, 2018.

\bibitem{Warden2018}
P.~Warden, ``Speech commands: A dataset for limited-vocabulary speech
  recognition,'' \emph{arXiv preprint arXiv:1804.03209}, 2018.

\bibitem{shan2018attention}
C.~Shan, J.~Zhang, Y.~Wang, and L.~Xie, ``Attention-based end-to-end models for
  small-footprint keyword spotting,'' \emph{Proc. Interspeech 2018}, pp.
  2037--2041, 2018.

\bibitem{Prabhavalkar2015}
R.~Prabhavalkar, R.~Alvarez, C.~Parada, P.~Nakkiran, and T.~N. Sainath,
  ``Automatic gain control and multi-style training for robust small-footprint
  keyword spotting with deep neural networks,'' pp. 4704--4708, 2015.

\bibitem{sainath2015convolutional}
T.~Sainath and C.~Parada, ``Convolutional neural networks for small-footprint
  keyword spotting,'' in \emph{Proceedings of Interspeech}, 2015.

\bibitem{Yu2018}
M.~Yu, X.~Ji, Y.~Gao, L.~Chen, J.~Chen, J.~Zheng, D.~Su, and D.~Yu,
  ``Text-dependent speech enhancement for small-footprint robust keyword
  detection,'' \emph{Proc. Interspeech 2018}, pp. 2613--2617, 2018.

\bibitem{wang2018supervised}
D.~Wang and J.~Chen, ``Supervised speech separation based on deep learning: An
  overview,'' \emph{IEEE/ACM Transactions on Audio, Speech, and Language
  Processing}, vol.~26, no.~10, pp. 1702--1726, 2018.

\bibitem{du2014robust}
J.~Du, Q.~Wang, T.~Gao, Y.~Xu, L.-R. Dai, and C.-H. Lee, ``Robust speech
  recognition with speech enhanced deep neural networks,'' in \emph{Fifteenth
  Annual Conference of the International Speech Communication Association},
  2014.

\bibitem{seltzer2013investigation}
M.~L. Seltzer, D.~Yu, and Y.~Wang, ``An investigation of deep neural networks
  for noise robust speech recognition,'' in \emph{2013 IEEE international
  conference on acoustics, speech and signal processing}.\hskip 1em plus 0.5em
  minus 0.4em\relax IEEE, 2013, pp. 7398--7402.

\bibitem{narayanan2013ideal}
A.~Narayanan and D.~Wang, ``Ideal ratio mask estimation using deep neural
  networks for robust speech recognition,'' in \emph{2013 IEEE International
  Conference on Acoustics, Speech and Signal Processing}.\hskip 1em plus 0.5em
  minus 0.4em\relax IEEE, 2013, pp. 7092--7096.

\bibitem{wang2014training}
Y.~Wang, A.~Narayanan, and D.~Wang, ``On training targets for supervised speech
  separation,'' \emph{IEEE/ACM Transactions on Audio, Speech, and Language
  Processing}, vol.~22, no.~12, pp. 1849--1858, 2014.

\bibitem{wang2016joint}
Z.-Q. Wang and D.~Wang, ``A joint training framework for robust automatic
  speech recognition,'' \emph{IEEE/ACM Transactions on Audio, Speech, and
  Language Processing}, vol.~24, no.~4, pp. 796--806, 2016.

\bibitem{ioffe2015batch}
S.~Ioffe and C.~Szegedy, ``Batch normalization: Accelerating deep network
  training by reducing internal covariate shift,'' \emph{arXiv preprint
  arXiv:1502.03167}, 2015.

\bibitem{nair2010rectified}
V.~Nair and G.~E. Hinton, ``Rectified linear units improve restricted boltzmann
  machines,'' in \emph{ICML}, 2010, pp. 807--814.

\bibitem{tan2018convolutional}
K.~Tan and D.~Wang, ``A convolutional recurrent neural network for real-time
  speech enhancement,'' in \emph{Proceedings of Interspeech}, 2018, pp.
  3229--3233.

\bibitem{sainath2013learning}
T.~N. Sainath, B.~Kingsbury, A.-r. Mohamed, and B.~Ramabhadran, ``Learning
  filter banks within a deep neural network framework,'' in \emph{2013 IEEE
  Workshop on Automatic Speech Recognition and Understanding}.\hskip 1em plus
  0.5em minus 0.4em\relax IEEE, 2013, pp. 297--302.

\bibitem{Tang2017}
R.~Tang and J.~Lin, ``Honk: A pytorch reimplementation of convolutional neural
  networks for keyword spotting,'' \emph{arXiv preprint arXiv:1710.06554},
  2017.

\bibitem{Kingma2014}
D.~P. Kingma and J.~Ba, ``Adam: A method for stochastic optimization,''
  \emph{arXiv preprint arXiv:1412.6980}, 2014.

\end{thebibliography}

% \begin{thebibliography}{9}
% \bibitem[1]{Davis80-COP}
%   Pete Warden,
%   ``Speech Commands: A Dataset for Limited-Vocabulary Speech Recognition,''
%   \arXiv:1804.03209, 2018.
% \bibitem[2]{Rabiner89-ATO}
%   L.\ R.\ Rabiner,
%   ``A tutorial on hidden Markov models and selected applications in speech recognition,''
%   \textit{Proceedings of the IEEE}, vol.~77, no.~2, pp.~257-286, 1989.
% \bibitem[3]{Hastie09-TEO}
%   T.\ Hastie, R.\ Tibshirani, and J.\ Friedman,
%   \textit{The Elements of Statistical Learning -- Data Mining, Inference, and Prediction}.
%   New York: Springer, 2009.
% \bibitem[4]{YourName17-XXX}
%   F.\ Lastname1, F.\ Lastname2, and F.\ Lastname3,
%   ``Title of your INTERSPEECH 2019 publication,''
%   in \textit{Interspeech 2019 -- 20\textsuperscript{th} Annual Conference of the International Speech Communication Association, September 15-19, Graz, Austria, Proceedings, Proceedings}, 2019, pp.~100--104.
% \end{thebibliography}

\end{document}